\documentclass[12pt,preprint]{aastex}
\input epsf.tex
\newcommand{\be}{\begin{equation} }
\newcommand{\ee}{\end{equation}}
\begin{document}
\title{Helium-Core White Dwarfs in Globular Clusters}
\author{Brad M.\ S.\ Hansen$^{1,2}$, Vassiliki Kalogera$^{3}$, and Frederic A.\ Rasio$^{3}$}
\affil{}
\affil{$^1$ Hubble Fellow, Division of Astronomy, University of California, 8971 Math Sciences, Los Angeles, CA, 90095}
\affil{$^2$ Department of Astrophysical Sciences, Princeton University, Princeton, NJ, 08544}
\affil{$^3$ Department of Physics \& Astronomy, Northwestern University, 2145 Sheridan Rd, IL, 60208}

\begin{abstract}
We examine the theoretical implications of a population of low-mass helium-core 
white dwarfs in globular clusters. In particular, we focus on the observed population 
in the core of NGC~6397, where several low-mass white dwarf canditates have been 
identified as ``non-flickerers'' by Cool and collaborators. Age and mass estimates 
from cooling models, combined with dynamical and evolutionary
 considerations, lead us to infer that the dark binary companions are C/O white dwarfs rather than
neutron stars. Furthermore, we find that the progenitor binaries very likely underwent an 
exchange interaction within
the last $10^9$~years. We examine the prospects for detecting a similar population in other globular
clusters, with particular attention to the case of 47~Tuc.
\end{abstract}


\section{Introduction}

Binaries play a central role in the internal dynamical processes that drive globular cluster
evolution (e.g., Hut et al.\ 1992; Bailyn 1995).
 Of particular
importance is the primordial binary star population (Heggie 1975; Hut et al.\ 1992), 
which provides a crucial internal energy source for the cluster by virtue of inelastic scattering 
encounters.  Some of the more dramatic products of these encounters include recycled pulsars 
(Rappaport, Putney, \& Verbunt 1989; Phinney 1996), low-mass X-ray binaries (Verbunt \& Johnston 1996), 
cataclysmic variables (Di Stefano \& Rappaport 1994; Cool et al. 1995; Grindlay et al. 2001) and blue stragglers 
(Sigurdsson, Davies, \& Bolte 1994; Lombardi, Rasio, \& Shapiro 1996).

Cool et al.\ (1998) and Edmonds et al.\ (1999) (hereafter CGC and EGC, respectively) have
reported the detection of a new stellar population near the center of the core-collapsed
globular cluster NGC~6397. They
named them ``non-flickerers'' and tentatively identified them as low-mass helium-core white
dwarfs (hereafter HeWDs). Several similar, but fainter, objects have been 
discovered by Taylor et al. (2001) in the same cluster. HeWDs are the result of 
the truncated evolution of
low- and intermediate-mass stars in binaries (Kippenhahn, Kohl \& Weigert 1967), in which the hydrogen
envelope of an evolving star is removed before the degenerate core is massive enough to burn
helium to carbon. Such objects are a generic by-product of mass transfer in close binaries with
evolved low-mass donor stars and may offer insight into the formation history of the compact object
population and its coupling to the global cluster evolution.

In this paper we use white dwarf cooling models, together with simple binary evolution and
dynamical models, to examine the nature of the ``non-flickerers,'' 
extending the initial discussion of EGC. We confirm that they must be HeWDs
with more massive, dark binary companions.
We place these systems in the broader dynamical context appropriate
to the underlying binary population and examine the results as a function of globular cluster parameters, 
with particular application to NGC~6397 and 47~Tuc.

The observations and their immediate implications are briefly summarized in \S2. In \S 3 we discuss
the formation and cooling of HeWDs. Applications to NGC~6397 and 47~Tuc
are presented in \S4 and \S5, respectively, while other clusters are discussed briefly in \S6.

\section{The `Non-flickerers'}

\label{disc}

The ``non-flickerers'' (NFs) were discovered by CGC during a search for cataclysmic variables (CVs)
in the core of NGC~6397 (Cool et al.\ 1995; Grindlay et al.\ 1995). 
 In the UV, CVs lie between the white dwarf sequence and the main sequence
by virtue of emission from the inner parts of the accretion disk. In the redder optical bandpasses, the
disk is fainter and the donor star becomes more apparent, so that the system is found closer to the main sequence.
The NFs differ from the CVs in that they show no evidence for a red donor star and, 
furthermore, they show none of the characteristic
 UV ``flicker'' associated with the CV accretion disk. On the basis of this, CGC coined the term
``non-flickerers.''
They advanced the hypothesis that the NFs were hot HeWDs, which lie above the
traditional white dwarf sequence by virtue of their smaller masses and thus larger radii. 
EGC presented a
spectrum for one of the NFs and derived temperature and gravity constraints 
( $\log g=6.25 \pm 1.0$, $T_{\rm eff} = 17500\pm 5000\,$K)
consistent with models of low-mass white dwarfs 
(Hansen \& Phinney 1998a; Benvenuto \& Althaus 1999). 

The masses inferred for these white dwarfs (see section~\S~\ref{models}) are in the range 
$0.2-0.25 M_{\odot}$, which is considerably smaller than the main sequence
turnoff mass ($\sim 0.8 M_{\odot}$). If these white dwarfs were single they would tend to leave
the core of the cluster on a timescale comparable to the central two-body relaxation time
(Fregeau et al.\ 2002).
The relaxation time in the core of NGC~6397 is  {
\be
 \tau_{\rm rc} \simeq 8 \times 10^5 {\rm yrs} 
 \left( \frac{ \sigma}{4.5 \rm km.s^{-1}} \right)^3 \left( \frac{M_{\rm wd}}{0.3 \rm M_{\odot}} \right)^{-1}
 \left( \frac{M_{\rm to}}{0.8 \rm M_{\odot}} \right)^{-1} \left( \frac{n}{1.5 \times 10^6 \rm pc^{-3}} \right)^{-1}  
\ee
}
 where we have used the central luminosity density from Djorgovski (1993), and we assumed a 
mass-to-light ratio of~3 and a (conservative) average stellar mass of $1 \rm \,M_{\odot}$ since 
the core is likely dominated by heavy stellar remnants (King, Sosin \& Cool 1995). For comparison
the central relaxation time estimated
by Djorgovski (1993) is $\simeq 8 \times 10^4$~years.  

On timescales longer than $\tau_{\rm rc}$, isolated low-mass white dwarfs should have
diffused out of the core as mass segregation develops in the cluster (Fregeau et al.\ 2002).
Yet, the NF positions measured by CGC are strongly concentrated towards the cluster centre. 
This clearly suggests that the NFs must be members of
binary systems with total masses comparable to or greater than the main sequence 
turnoff mass.  
Indeed, the binary companions could have been expected, since low-mass white 
dwarfs are formed only by
truncated stellar evolution in close binaries\footnote{Han, Podsiadlowski \& Eggleton (1994) 
note that HeWDs can potentially be formed
by the evolution of isolated low-mass ($<1\rm \,M_{\odot}$) stars. However, this formation
path can only occur for population~I stars and is therefore not relevant to globular
clusters. In addition, it appears to produce white dwarfs of mass $\simeq 0.4\rm \,M_{\odot}$, 
i.e., too massive to describe the NFs in any event.}.  A companion of
sufficient mass to explain the central location ($\gtrsim 0.6 \rm M_{\odot}$) would
be of comparable or greater luminosity and much redder than the NF.
Thus, the total binary mass must be dominated by a dark component, such as
a neutron star (as suggested by CGC) or a C/O white dwarf.
This is also consistent with
the observation of a Doppler shift of about $200\,{\rm km}\,{\rm s}^{-1}$ 
(interpreted as an orbital velocity) for one of the NF by EGC.

Since we will discuss the role of exchange interactions and the resultant recoils, we also note the half-mass relaxation time is $\sim 2 \times 10^8$ yrs
for NGC~6397.
On this timescale massive bodies kicked out of the core (such as hard binaries), will segregate back to the centre. Thus, for systems older than this, we
expect them to be concentrated towards the core, whereas younger systems could potentially be found in the outer parts. 


\section{Helium Core White Dwarfs}
\subsection{Origins}
\label{origin}
Low-mass white dwarfs are the result of binary stellar evolution. In the field
they are found as companions to millisecond pulsars (Ryba \& Taylor 1991;
 Phinney \& Kulkarni 1994;
Hansen \& Phinney 1998b) and in double degenerate systems (Bragaglia et al.\ 1990; Marsh, Dhillon, \& Duck 1995; 
Saffer, Livio, \& Yungelson 1998).
In binaries with initial separations $<200 \rm R_{\odot}$, a low-mass star overflows its Roche lobe 
as it evolves off the main sequence and
up the giant branch, but before core-helium ignition starts. The consequent mass loss truncates 
the stellar evolution and leaves the
remnant degenerate core to cool as a HeWD. The final binary 
configuration depends on the stability of
the mass transfer process and, in particular, the mass of the companion star.

If the accretor is a neutron star (for which we assume a mass of about $1.4\rm \,M_{\odot}$), then 
the mass transfer is stable as long as the donor mass is less than about $1.5\rm \,M_{\odot}$.  
This includes the stabilising effect of non-conservative mass transfer (the critical donor 
mass for dynamical instability decreases to about $1\rm \,M_{\odot}$
for conservative mass transfer; see Kalogera \& Webbink 1996). Mass transfer is driven by 
the radial expansion of the evolving red giant and the orbit expands
until the donor envelope mass is exhausted. The radius of the red giant is primarily 
determined by its core mass, and consequently the final orbital
period should correlate with the white dwarf mass (see, e.g., Rappaport et al.\ 1995). 
If instead the accretor is a C/O white dwarf, then there is little possibility
of stable mass transfer for donors with mass $M>0.8\rm \,M_{\odot}$. Indeed Han (1998) and Nelemans 
et al.\ (2000) have investigated the formation of double degenerate systems
and find that all systems undergo a common-envelope phase during the second mass 
transfer episode, which results in the formation of the low-mass
HeWD\footnote{The {\em first\/} mass transfer episode, resulting in 
the formation of the more massive white dwarf, may be stable or may also
lead to a common envelope.}. This difference in behaviour is a consequence of the 
larger mass of the neutron star and the tendency for the orbit to expand
when mass is transferred from a lighter donor to a heavier accretor.

The above considerations apply to isolated binaries and so describe the behaviour of the 
primordial binary population. 
However, in a dense globular cluster, there are more important {\em dynamical\/} paths 
that can lead to the production of HeWDs. 
Of particular importance are exchange interactions involving hard binaries (those 
with binding energies exceeding the typical kinetic energy of other cluster stars).
In most cases it is the least massive of the three stars involved in a strong interaction
that will be ejected. This both promotes the formation of HeWDs (by increasing the average 
secondary mass in binaries with a neutron star
or C/O white dwarf primary -- thereby increasing the likelihood that there will be 
a mass transfer episode within a Hubble
time) and contributes to their eventual removal from the binary (since the HeWD mass 
$\sim 0.1-0.4\rm M_{\odot}$ is significantly smaller than the masses of typical main-sequence 
intruders). Although they owe their provenance to dynamical interactions, the products of 
post-exchange mass-transfer binaries that go through stable mass transfer are
still subject to the same basic stellar evolution as primordial binaries. Thus, they should 
follow the same orbital period -- companion mass relation as the primordial systems. 

We must also consider
those evolutionary pathways in which the dynamical interactions lead directly to close 
degenerate binaries. These include tidal captures followed by mergers (Ray, Antia \& Khembavi 1989;
McMillan, Taam, \& McDermott 1990; Rasio \& Shapiro 1991) or direct
collisions between red giants and compact objects (Verbunt 1987; Bailyn 1988; Rasio \& Shapiro 1991;
Davies, Benz, \& Hills 1991). In all cases the resulting merger
will leave the red giant core (or proto-white-dwarf) and the original compact object in a 
binary, with most of the giant envelope expelled to infinity. The cross section for these
processes in globular clusters is dominated by gravitational focussing
and thus proportional to the red giant radius.
We may thus determine the distribution of final white dwarf masses
by calculating the fraction of the total collision cross section $\Sigma$ associated 
with stages before the red giant has evolved to the point where the
core mass $M_{\rm c}$ has a particular value $M$, i.e., 
\be
\frac{\Sigma (M_{\rm c}<M)}{\Sigma (M_{\rm c}<M_{\rm f})} = \frac{ \int_0^{M} R(M_{\rm c}(t)) dt}{\int_0^{M_{\rm f}} R(M_{\rm c}(t)) dt}
\ee
where $M_{\rm f}$ is the core mass at the point of core helium ignition.
The result is shown in Figure~\ref{dsdm}, where we have used the models of Hurley, 
Pols, \& Tout (2000). We see that 75\% of the 
final products have $M_{\rm wd}<0.3\, {\rm M_{\odot}}$, a result similar to that of 
Verbunt (1988). However, the resulting binaries
are not particularly hard, as the relative kinetic energy of the two stars is similar 
to the binding energy of the giant envelope and there is little inspiral associated with 
the ejection of the envelope (Rasio \& Shapiro 1991).
The final orbital separations are therefore comparable to the original giant radius 
and these binaries will have
lifetimes against further interactions that are similar to those of systems 
undergoing stable mass transfer.


\vspace{2mm}

\subsection{ Cooling Models}
\label{models}

Low-mass white dwarfs fall between the traditional white-dwarf cooling sequence and the main sequence because the 
non-relativistic degenerate equation of state implies $R \sim M^{-1/3}$, making lower mass dwarfs cooler at fixed
luminosity. For relatively young (\& thus hot) white dwarfs, there is
the additional feature that a small mass fraction of hydrogen on the surface can increase the radius further. Thus, the CMD positions
of the NFs  argue in favour of their low mass. 

Low-mass white dwarfs also have helium cores, rather than carbon/oxygen cores, because their prior evolution was truncated during
the mass transfer phase and they never reached the core helium burning phase. This affects the cooling rate because the heat capacity
per unit mass of the white dwarf is thereby increased by a factor $\sim 3$. Thus, HeWDs will be 
brighter than contemporaneous C/O dwarfs due to their slower cooling. There may also be a contribution from residual hydrogen burning if these low mass
white dwarfs possess thick hydrogen surface layers. This effect was first noted by Webbink (1975) and results from the
fact that lower mass dwarfs (with lower surface gravities) also have smaller pressures at the base of the envelope and thus weaker CNO-cycle
shell burning and fewer thermal pulses.
 This trend has been investigated in
recent years (Alberts et al.\ 1996; Driebe et al.\ 1999; Serenelli et al.\ 2001) as a potential explanation for the puzzling discrepancy
between the timing age of the PSR~J1012+5307 and the cooling age of the white dwarf companion if it had only a moderate surface
layer (Lorimer et al.\ 1995; Hansen \& Phinney 1998b). On the other hand, the limits on the companion to the PSR~B1855+09 (van Kerkwijk et al.\ 2000) are
inconsistent with a thick layer mass ($M_{\rm H}<3 \times 10^{-4} \rm M_{\odot}$) but consistent with the moderate layer cooling age derived in Hansen \& Phinney (1998b). 
This dichotomy is probably because the derived layer masses are sensitive to some aspects of the physics like the treatment of diffusion
of CNO elements and the mass
loss prescriptions. 



The hydrogen layer thickness is also affected by the evolutionary history of the white dwarf. The various papers
advocating large hydrogen layer masses (Driebe et al.\ 1995; Serenelli et al.\ 2001) calculated the conditions relevant
for stable mass transfer driven by Roche-lobe overflow, since they are intended to describe the companions to field
millisecond radio pulsars. Thus, while these models suffice to describe the companions to field neutron stars, they do not
necessarily apply to the post common-envelope systems i.e. those in which the primary is a C/O white dwarf.

To estimate the remnant layer mass in the case of a common-envelope episode, we note that the inspiral takes place on timescales much
shorter than the envelope thermal time, so that the heat deposition is considerably more rapid than in the stable
mass transfer case; i.e., the thermal structure of the envelope has little chance to respond to the inspiral.
 Thus, we assume that only that fraction of the original envelope mass that lies
within the final Roche lobe will fall back onto the system\footnote{This is a conservative assumption as any expansion in
response to the dissipation of orbital energy will only reduce the fallback mass still further.}, i.e.,
$ M_{\rm f}/M_{\rm 0} \sim (a_{\rm f}/a_{\rm 0})^{3-\alpha}$, where $M_{\rm f}$ and $M_{\rm 0}$ are the final and initial envelope masses and 
$\alpha$ is the power law slope of the envelope density profile ($\alpha$=1.5 for
an isentropic ideal gas). The ratio of pre- and post-inspiral separations varies from $\sim$25--150; i.e.,
$10^{-4}$--$10^{-2}$ of the original envelope mass is retained. Furthermore, the fallback mass is subject to
competitive accretion by
{\em both} bodies in the binary, so that the HeWD is likely to accrete only that fraction
corresponding to its relative fraction of the total Roche lobe volume i.e., $\sim M_{\rm wd}/M_{\rm total} \sim \frac{1}{3}-\frac{1}{2}$.
We use the population synthesis models of Rasio, Pfahl \& Rappaport (2000) \& Rappaport et al.\ (2001) to calculate the
post-exchange binary population and estimate
the hydrogen layer masses as above. The results are shown in Figure~\ref{mh}.
 We see that NFs with C/O white dwarf companions
have considerably smaller hydrogen layer masses than is found for the stable-transfer neutron star case or for
the case of a common-envelope system containing a neutron star. This arises from two effects. The first is that the C/O white
dwarf is less massive than the neutron star and thus spirals in further to release a fixed amount of binding energy.
The second is that binaries containing C/O white dwarfs can enter the common envelope stage with much lower mass
secondaries than in the neutron star case (see \S\ref{origin}), so that the envelopes that must be expelled are less massive to begin with.



To illustrate the effect of this range of hydrogen layer masses, we use
the code of Hansen \& Phinney (1998a), which incorporates detailed interior and atmospheric physics and 
an extended range of hydrogen layer masses
 as required by 
this study.

Figure~\ref{LowH} shows the cooling of HeWD with a ``moderate'' hydrogen surface layer mass
fraction of
 $q_{\rm H}=10^{-4}$. For such a layer mass, surface hydrogen burning makes a negligible contribution
to the luminosity.
 Also shown is the cooling curve for a standard 0.5~$\rm M_{\odot}$ carbon/oxygen
core white dwarf and for a binary containing two identical $0.5 \rm M_{\odot}$ white dwarfs. This latter curve
shows the extent to which normal white dwarf binaries can be confused with low-mass objects. The original
CGC white dwarfs are thus in the age range $10^7-10^8$ years in this model, while the later Taylor et al.\ (2001)
 dwarfs are $\sim 1$~Gyr old.
All the white dwarfs have masses $\sim 0.17-0.21 M_{\odot}$. Also shown is the optical companion to the millisecond pulsar
in 47~Tuc, detected by Edmonds et al.\ (2001). This object is of even lower mass ($\sim \rm 0.14 M_{\odot}$) with an age
$\sim 7 \times 10^7$~yrs.

Figure~\ref{HiH} shows the same cooling behaviour, but now for an initial surface hydrogen layer mass fraction of $q_{\rm H}=0.03$. 
This corresponds to layer masses comparable with those of Driebe et al.\ (1999). The strong influence of residual hydrogen
burning is now clearly seen, with the lowest mass dwarfs showing a prolonged lifetime at high luminosities. For more
massive objects (M$>0.25 \rm M_{\odot}$) this effect gets weaker, since the nuclear burning (driven by the higher pressures
at the base of the hydrogen envelope) is stronger and the available fuel is consumed on timescales shorter than
a Hubble time. This is consistent with the observations of the PSR1855+09 companion, whose mass ($\sim 0.23 \rm M_{\odot}$)
is large enough to reduce the retardation due to hydrogen burning. In this model, the CGC dwarfs are again $\sim 0.2 \rm M_{\odot}$,
but now their ages range from 0.1-3~Gyr. Furthermore, the Taylor et al.\ (2001) objects are then very old, ( $\sim 15$~Gyr).
The Edmonds et al.\ (2001) object has a slightly larger mass than before ($\sim 0.18 \rm M_{\odot}$), but still an age
$\sim 10^8$ yrs.



Thus, to summarise, it appears that there are two possibilities for the nature of
the bright NFs in NGC~6397.
The first is that they are relatively
young (ages $<10^8\,$yr) with moderate surface hydrogen envelopes, in which residual hydrogen burning plays no part
in stalling the cooling evolution. The second possibility is that they are older (ages $>10^9\,$yr) but with
thicker hydrogen envelopes, in which much of the luminosity is supplied by hydrogen burning. In the next
section, we investigate these
two possibilities further, taking into account other factors such as the dynamical history of the cluster.

\section{ Application to NGC~6397}

\label{His1}

\subsection{Young NF}
In the first scenario, the relative youth of the HeWDs implies that their progenitors evolved off the
main sequence within the last $\sim 10^8\,$yr, i.e., the progenitor mass was not much larger than the
current turnoff mass (about $0.8\rm \,M_{\odot}$). Thus, if the dark companion is a $1.4\rm \,M_{\odot}$ neutron 
star, the donor/accretor
mass ratio is $<0.6$ and the mass transfer is stable. The binary orbit will then expand during mass transfer, 
and we expect the final system to obey the usual companion mass -- orbital period relation. In a dense, 
core-collapsed cluster such as
NGC~6397, it is very likely that the white dwarf will be removed from such a wide binary during a later
interaction. To examine this more quantitatively, consider the approximate lifetime of the binary
to exchange interactions (e.g., Davies 1995) {
\be
 \tau_{\rm ex} \sim 10^{11}\,{\rm yr} \left(\frac{\sigma}{10\,{\rm km}\,{\rm s}^{-1}}\right) 
\left( \frac{n}{10^5\,{\rm pc}^{-3}} \right)^{-1} \frac{1}{a} \frac{1}{M_1+M_2+M_3
} \label{Tdis}
\ee
}
 where $M_1$, $M_2$, and $M_3$ are the masses (in $\rm M_{\odot}$) of the two binary components and 
of the incoming perturber, respectively, and $a$ is the binary
semi-major axis (in $\rm R_{\odot}$).
Converting $a$ to binary orbital period, assuming $M_1+M_2\simeq 2\rm \,M_{\odot}$, $M_3=0.8\rm \,M_{\odot}$, 
and adopting the
central luminosity density of Djorgovski (1993) with a mass-to-light ratio of 3 and the central velocity 
dispersion of Pryor \& Meylan (1993), we infer the maximum orbital period that a surviving binary is 
likely to have for a given lifetime to exchange {
\be
P_{\rm orb} \sim 3\,{\rm d}\, \left( \frac{\tau_{\rm ex}}{10^8\,{\rm yr}} \right)^{-3/2} 
\left( \frac{\sigma}{4.5\,{\rm km}\,{\rm s}^{-1}} \right)^{3/2} 
\left( \frac{n}{1.5 \times 10^6\,{\rm pc}^{-3} } \right)^{-3/2} \label{Porb}.
\ee
}
We have chosen the exchange time to be a nominal lifetime for the NFs. We can furthermore convert 
this into a relation between
white dwarf mass and exchange lifetime, using the orbital period -- core mass relation of Rappaport 
et al.\ (1995). This yields a range of allowed white dwarf masses
{
\be
 M_{\rm wd} < 0.2 \rm M_{\odot} \left( \frac{\tau_{ex}}{10^8\,{\rm yr}} \right)^{-6/25} 
\left( \frac{\sigma}{4.5\,{\rm km}\,{\rm s}^{-1}} \right)^{6/25} 
\left( \frac{n}{1.5 \times 10^6\,{\rm pc}^{-3}} \right)^{-6/25} \label{Mwd}
\ee
}
The half-mass relaxation time for NGC~6397 $\tau_{\rm rh}\simeq 2 \times 10^8\,$yr, so this estimate 
is not sensitive to whether
or not post-exchange systems acquire significant recoil velocities. Those that do will diffuse 
back into the core on the timescale of interest. 
These considerations are also consistent
with the Doppler velocity of about $200\,{\rm km}\,{\rm s}^{-1}$ measured by Edmonds et al.\ (2001), 
as {
\be
 P_{\rm orb} = 2 \pi \frac{G M_{\rm tot}}{V^3} \sim 2\,{\rm d} \left( \frac{M_{\rm tot}}{2\rm \,M_{\odot}} \right) 
\left( \frac{V}{200\,{\rm km}\,{\rm s}^{-1}} \right)^{-1}.
\ee
}
However, the white dwarf age is an underestimate in this case because
systems with final orbital periods $<10\,$d are the result of a competition between nuclear 
evolution and magnetic braking (Pylyser \& Savonije 1988; Ergma 1996;
Podsiadlowski et al.\ 2002), a process which takes $>10^9\,$yr. As an example, 
models B20,C20 and D20 of Pylyser \& Savonije (the most
appropriate to the probable progenitor systems) spend 1--4~Gyr in the mass transfer phase, 
10--20 times the probable lifetime to exchange. Furthermore, the
stable mass transfer evolutionary models appropriate to this phase (Driebe et al.\ 1999; 
Serenelli et al.\ 2001) predict the larger hydrogen layer masses
more appropriate to the `old NF' scenario examined in the next section.  Thus, it is 
difficult to make a consistent cooling and dynamical model for the NF if
we assume the dark companion to be a neutron star.

The essential problem is that stable mass transfer leads to post-transfer binaries that are too wide.
More compact systems that avoid further exchange interactions can be formed only 
if the system undergoes common-envelope evolution. This can happen if the dark companion mass 
is less than the progenitor mass, i.e.,
if the accretor is a normal C/O white dwarf rather than a neutron star. It also allows for arbitrarily 
young NFs, since the average
C/O white dwarf mass is $0.5-0.6 \rm \,M_{\odot}$, less than the current turnoff mass. The resultant 
common-envelope inspiral makes the binaries hard enough that they are no longer susceptible to 
further exchanges. Furthermore, our estimates of the hydrogen layer
mass in section~\ref{models} are also consistent with the values required to give the young 
model ages for the observed systems.
Thus, the assumption of a C/O white dwarf companion results in a consistent solution, with 
both dynamical lifetimes and
cooling ages within the observational constraints.

\subsection{Old NF}

The above arguments were built on models in which the hydrogen layer mass was small and the 
NFs were consequently young. If the NFs had larger hydrogen layer masses, their ages could be 
larger, of order several Gyr. This is the
type of white dwarf expected from evolutionary models of stable mass transfer in close neutron star binaries. 
However, in order for the white dwarf to survive for the estimated cooling time $\sim 3 \times 10^9\,$~yr,
equation~(\ref{Mwd}) predicts $M_{\rm wd} <0.09 \rm M_{\odot}$.

The only way that the dark companion could be a neutron star is if the progenitor mass were large 
enough that the mass transfer was unstable and the system entered a common envelope, i.e., if the 
progenitor were sufficiently more massive
than the neutron star. For a progenitor mass $>1.4 \rm M_{\odot}$ and a cluster age $\simeq 12\,$Gyr, a 
white dwarf age $>5\,$Gyr is required. Although our inferred NF ages are not that large, we have some 
freedom to tune the white dwarf age by increasing the hydrogen layer mass still further. However, other 
factors come into play, which we now examine.

Very tight binaries will not always survive in the dense cluster core. If the system is too compact, 
gravitational radiation will merge the binary anyway (thereby destroying the NF).
 The timescale for this is {
\be
\tau_{\rm GR} \sim 3 \times 10^8\,{\rm yr} \left( \frac{a}{R_{\odot}} \right)^4 
\left( \frac{M_{\rm tot}}{1.6\rm \,M_{\odot}} \right)^{-1}
\left( \frac{ 1.4\rm \,M_{\odot}}{M_{\rm ns}} \right) 
\left( \frac{ 0.2\rm \,M_{\odot}}{M_2} \right) \label{Tmerge}
\ee
}

Thus, by equating (\ref{Tdis}) and (\ref{Tmerge}), we find the maximum age to which the 
white dwarf-neutron star binary can survive in the
core of NGC~6397 (see Figure~\ref{Ta})
\be
\tau_{\rm max} \sim 1\,{\rm Gyr} \left( \frac{M_2}{0.2\rm \,M_{\odot}} \right)^{-1/5}. \label{Tmax}
\ee
We see that this clearly rules out the old NF solution, as the combination of inferred ages 
and masses would violate equation~(\ref{Tmax}).


\subsection{Fainter NF}

Taylor et al.\ (2001) detected a second group of potential NFs in NGC~6397 at fainter magnitudes. 
In the ``young~NF'' scenario, these objects
have ages $\sim 10^9\,$yr and are just consistent with equation~(\ref{Tmax}). 
In the ``old~NF'' scenario, these are $>10^{10}\,$yr
old and violate all the constraints. Thus, if these objects share a common origin with the brighter NFs, 
they are also consistent with the ``young~NF'' solution. 

We mention this group separately because an alternative explanation has been suggested for them 
by Townsley \& Bildsten (2002):
that they could be old, low-luminosity CVs. However, it is important to note that this explanation 
cannot be applied to the
brighter NF objects discussed by CGC.



\subsection{Observational Implications}

Given the above considerations, we favour the following model for the observed NF population in NGC~6397. 
They are HeWD
 with moderate ($\sim 10^{-4} \rm M_{\odot}$) hydrogen surface layers and ages $\sim 10^8$ years (for the
bright population). They have formed
from progenitors of mass $\sim 0.8 \rm M_{\odot}$ in binaries with C/O white dwarfs, and have undergone common envelope evolution.

Our conclusion that there is no self-consistent and convincing scenario that can allow for neutron star
companions has an important observational consequence:
none of the NFs should show the X-ray signatures characteristic of millisecond pulsars. 
Grindlay et al.\ (2002) have examined the core of NGC~6397 with the {\em Chandra\/} satellite, 
identifying many sources (including a millisecond pulsar and
several potential CVs), none of which are coincident with the NFs. This ``surprising'' (in their words) 
finding is consistent with our analysis.

One could also ask about the possibilities for finding fainter NFs in deeper searches. Here the maximum 
age constraint (eq.~\ref{Tmax}) 
implies that we do not expect to find any NFs
fainter than the second population observed by Taylor et al.\ (2001).

\subsection{Surviving Core Collapse}

So far we have treated the NGC~6397 core as one of constant density and velocity dispersion. However, 
NGC~6397 is a ``core-collapsed''
cluster, where the very definition of a core is somewhat uncertain. We have used the central density 
and velocity dispersion from Djorgovski (1993) \& Pryor \& Meylan (1993) respectively, although the 
true density may be lower (and thus lifetimes longer; A.\ Cool, personal communication). On the other 
hand, the central densities previously reached at the onset of core collapse were potentially even 
higher and the very low binary fraction (Bolton, Cool, \& Anderson 1999) suggests that only the very
hard binaries have survived. As a result, we feel that our conclusion that the NFs are young, with C/O 
white dwarfs as companions, is independent of the details of the past cluster evolution.

Since the NFs are $\sim 10^8\,$yr old, the progenitor binaries must have either survived core collapse 
or been produced during the collapse. Within the ``young NF'' scenario we may constrain the progenitor 
configuration as follows. The donor mass
is required to be $\sim 0.8\rm \,M_{\odot}$ (since the observed NF masses are larger than the 
$\sim 0.13\rm \,M_{\odot}$ of the
initially hydrogen exhausted core of a turnoff star, i.e., the progenitor must have instigated 
mass transfer after leaving the main sequence)
 and the observed core mass tells us the giant radius at which point mass transfer started. Thus
we may use the Rappaport et al.\ (1995) relation to infer the {\em initial\/} orbital separation 
$a_0$ of a dynamically unstable system. We find {
\be
a_0 = \frac{0.5 \rm R_{\odot} + 2.36 \rm R_{\odot} \left( M_2/0.2 \rm M_{\odot} \right)^{4.5}}
{ 0.38 + 0.2 \log \left( 0.8{\rm  M_{\odot}}/M_1 \right)}
\ee
}
where $M_1$ and $M_2$ are the C/O white dwarf and HeWD masses, respectively. For $M_1 \simeq 0.5\rm \,M_{\odot}$ 
and $M_2 \simeq \rm 0.2\,M_{\odot}$,
this yields $a_0 \simeq 7\rm \,R_{\odot}$, corresponding to a period $P_{\rm orb} \simeq 1.8\,$d. The exchange 
lifetime for such a binary (using the
same parameters as before) is
$\sim 2 \times 10^8\,$yr,  similar to the current ages of the white dwarfs. This may also  
explain why we do not observe
HeWDs of greater mass (theoretically up to $0.4\rm \,M_{\odot}$; these more massive HeWDs are prevalent
among the field double degenerate systems. See Bergeron, Saffer \& Liebert 1992). The rapid increase in $a_0$ with $M_2$ 
leads to a rapid decrease in the lifetime to exchange.
For $M_2 \simeq 0.3\rm \,M_{\odot}$ we find $a_0 \simeq 36\rm \,R_{\odot}$ and the exchange time is now
only $4 \times 10^7\,$yr. Thus, in order
for a star to grow a helium core $>0.25\rm \,M_{\odot}$ and avoid Roche Lobe overflow 
(since that would lead to a CE and immediately halt the
evolution) it would have to be in such a wide binary that another exchange interaction would
quickly ensue. 
Thus, not only did the mass transfer episode occur only $\sim 10^8\,$yr ago, but the progenitor 
binary configuration had
a similar lifetime. We therefore expect the progenitor stars to have typically participated 
in several exchange interactions before forming the final
binary configuration that led to the mass transfer.

\subsection{An evolving population}

Since any given binary incarnation cannot survive much longer than
1~Gyr in the core of NGC~6397, compact objects such as white dwarfs and neutron stars may 
undergo several life-cycles with
different companions over the course of a Hubble time.
Not only do we include soft binaries in this statement but also neutron stars and massive white dwarfs with close
companions which undergo gravitational wave inspiral.  After disrupting their inspiralling companions  
they will be available again as single objects to exchange into other binaries.

This also implies that the cluster should contain binaries in which the neutron star 
was recycled by accretion in a previous incarnation,
resulting in a millisecond pulsar with a close main sequence companion. This is most likely 
the nature of the recently
discovered first radio millisecond pulsar in NGC~6397 (D'Amico et al.\ 2001), which shows 
a variable optical companion consistent
with a main sequence star (Ferraro et al.\ 2001a). D'Amico et al.\ (2001) note that this is 
potentially the result of an exchange
interaction and Ferraro et al.\ (2001a) further note that the candidate companion is, 
in fact, one of the BY Draconis candidates
discussed by Taylor et al.\ (2001).

\section{Application to 47 Tuc}
\label{His2}

We have thus far focussed on the NFs discovered in NGC~6397. However, there are now 
{\em Chandra\/} and {\em HST\/} observations of
many globular cluster cores. In what follows, we discuss our expectations for the (currently undetected) NF population in 47~Tuc. The attraction of this cluster
is the rich data set spanning X-ray (Grindlay et al.\ 2001), UV (Ferraro et al.\ 2001b; Knigge et al.\ 2001), optical (Gilliland et al.\ 2001),
and radio (Camilo et al.\ 2000)
wavelengths. 47~Tuc is also interesting in that it is a dense, pre-core collapse cluster, and somewhat more massive than NGC~6397.

The nominal central density of 47~Tuc is $1.5 \times 10^5 {\rm pc}^{-3}$, ten times smaller than the value we used for
NGC~6397. This has the important consequence that much wider binaries can survive for a reasonable length of time without
suffering an exchange interaction. If we rescale equation~(\ref{Tmax}) to the 47~Tuc parameters (and for neutron star binaries)\footnote{
Note that this expression does not include the orbital period-mass relation for stable systems i.e., it represents the
limit obtainable with common-envelope inspiral. The corresponding limits for stable mass transfer products will be discussed
below.}, we find
\be
\tau_{\rm max} = 10 \rm Gyr \left( \frac{M_2}{0.2 \rm M_{\odot}} \right)^{-0.2}. \label{tmax2}
\ee
Thus, the potential age range of NF binaries in 47~Tuc is much larger than for NGC~6397.



The radio observations find a large population of binary millisecond pulsars in 47~Tuc (Manchester et al.\ 1991; Camilo et al.\ 2000), many
of which appear to have low-mass white dwarf companions. In one case, 47~Tuc~U, the NF companion has been detected
optically (Edmonds et al.\ 2001), with a mass $\simeq 0.14\rm \, M_{\odot}$ and an age $\sim 10^7-10^8\, $yr, even for
a large hydrogen surface layer.
The lower density of 47~Tuc means that, in contrast to the situation in NGC~6397, this could be the 
product of a primordial binary as well as a dynamically-produced binary.
Furthermore, the X-ray and UV searches have uncovered a range of CV candidates, shown in  
Figures~\ref{Cool5} and \ref{Ferr}. The NF candidates are also expected to be found in this 
region of the CMD, as shown by the cooling curves
in the figures. An effort to determine what fraction of the CV candidates are actually NFs will elucidate the binary interaction history of this cluster.



The NF population of 47~Tuc is thus expected to have three separate components (in contrast to NGC~6397):
those with
origins in the primordial binary population, and those resulting from exchange interactions, both with 
neutron star and C/O white dwarf companions. The products of stable mass-transfer systems are still somewhat
limited, as equation~(\ref{Mwd}), converted to 47~Tuc conditions, yields a mass-dependant upper limit to the
cooling age of a white dwarf of given mass
\be
\tau < 2.7 {\rm Gyr} \left( \frac{M_{\rm wd}}{0.2 \rm M_{\odot}} \right)^{-25/6}.
\ee
Note that this is stronger than the limit given by (\ref{tmax2}) because the orbital-period mass relation restricts these
systems 
to wider orbits. Using the binary interaction Monte Carlo code of Rasio, Pfahl, \& Rappaport (2000; see also
Rappaport et al.\ 2001), we have also examined the NF population that emerges from exchange interactions.
Figure~\ref{MT} shows the NF mass as a function of age for NF with both neutron star and C/O white dwarf
companions. The NFs with C/O white dwarfs show a wide range of masses and ages. The population with
neutron star companions is bimodal. For systems resulting from stable mass transfer, we can expect to see
objects of mass $0.15-0.2\rm \,M_{\odot}$ for up to $\sim 3\,$Gyr. More massive NFs will also have neutron
star companions, but only if they have undergone common-envelope evolution. This requires them to have
had more massive progenitors and are thus correspondingly older (figure~\ref{MT}) and thus cooler.
We also note that these more massive objects will be fainter
regardless of the initial hydrogen layer mass, as they lie above the mass range 
where significant retardation of the
cooling is possible (\S~\ref{models}).



To conclude, there is great potential for studying the NF population in 47~Tuc. The observed bright millisecond pulsar
companion is consistent with a stable mass transfer episode from a $0.8\rm \,M_{\odot}$ giant. Many other low-mass 
white
dwarfs are expected, with the brighter population being dominated by systems with 
C/O white-dwarf companions and the fainter 
population dominated by systems with neutron star companions.

\section{Other Clusters}
\label{Others}

 There are, as yet, no
NF candidates in any other clusters. However, we can use the known millisecond pulsar binaries to roughly
anticipate the conditions in other clusters. Figure~\ref{TT} shows the known pulsar binaries in the
distribution of lifetime to exchange versus merger lifetime. We have split them up into four
groups, depending on cluster density. 47~Tuc occupies a class of its own by virtue of its large
number of detections.



What is clear from this diagram is that very few of the cluster millisecond pulsar binaries are likely to be due to an undisturbed primordial
configuration; to survive undisturbed for a Hubble time they would need to lie in the top right-hand-side quadrant. Thus, nearly all of these clusters may
be expected to possess a population of close binaries sculpted by exchange interactions. In figure~\ref{TT} we also show the simulations of Rappaport et
al. (2001) for the 47~Tuc parameters (which clearly explain the trend of the observed binaries).  The majority of the predicted systems lie in the top
left-hand quadrant. These are still relatively wide binaries, vulnerable to further exchanges.  The observed systems in this area include the NGC~6397 pulsar 
system, which is most likely a recent exchange capture of a main sequence star and the system NGC~6266A, which has very similar parameters. The binary at
the extreme upper left of the plot is the one in NGC~6441. where the large companion mass and significant eccentricity argues for a recent white dwarf
capture. The bottom right-hand quadrant is occupied by very compact binaries. The lack of systems in this area is most likely because the recoil
velocities of post-exchange binaries (Phinney \& Sigurdsson 1991) will eject such tight binaries from the cluster (see Figure~\ref{Ta}).

Location of objects on this diagram is based purely on their dynamical properties. The observation of a low
mass white dwarf (NF) allows us to obtain a quantitative measure of the age of the system dating from the
end of mass transfer and to thereby test the model that goes into the construction of a figure such as
figure~\ref{TT}. In the case of NGC~6397, all the objects have consistent model ages $< 1$~Gyr, in agreement with
the location of the NGC~6397 pulsar. The other clusters in class A are also ideally suited for the detection
of bright NF.

\section{Conclusions}
\label{conclusions}

We have examined the various possible evolutionary pathways for the formation of NF systems, identified as HeWD by CGC,
 in globular cluster cores. We have payed particular attention to the cluster NGC~6397 where these objects were first discovered. We find that
 most likely the NGC~6397 NFs are HeWD with CO core white dwarfs as their dark companions. This model satisfies all
constraints imposed on the lifetimes and masses of white dwarfs (inferred from the observations and the dynamical properties of the host cluster). The low
masses of CO white-dwarf companions (relative to neutron-star companions) allow recent common-envelope evolution that produces (i) orbits tight enough to
avoid disruption due to dynamical interactions, and (ii) young, bright HeWD to explain their position in the CMD. Neutron-star dark
companions would be too massive and would result either in stable mass transfer (producing systems that are too wide to survive exchange interactions for
a significant time in a dense cluster core), or in HeWD too old and too faint.

The common envelope episode also results in a moderate mass hydrogen envelope on the
white dwarf surface, leading to an estimate of the cooling age that is consistent with
the dynamical considerations. A massive hydrogen envelope, in which nuclear burning is
important, would imply that the observed systems are older than they are likely to
be, based on their probability of survival in the cluster core.

We have extended our analysis of NGC~6397 to other globular clusters to illustrate how
the production of NFs depends on cluster environment. In particular,
we have studied the potential NF population in 47~Tuc, which is both less dense and more massive than
NGC~6397, and for which a wealth of observational data is becoming available. In this cluster
surviving NFs with neutron star companions are possible, but they will be older and more massive
(and correspondingly fainter) than those with C/O white dwarf companions. This is again
the consequence of whether binaries can survive intact, otherwise being transformed by
exchange interactions or gravitational wave-induced mergers.

We anticipate that these models can be extended to other clusters as observational results accrue. The inferred white dwarf cooling ages will allow us to
place constraints on the exchange interaction history of clusters in a manner that is independent of other methods, such as the properties of millisecond
pulsar binaries.

\acknowledgments 
We thank Eric Pfahl and Saul Rappaport for many useful discussions and comments on
the manuscript.
BH acknowledges support from Hubble Fellowship HF-01120.01, which was provided by NASA through 
a grant from the Space Telescope Science Institute, 
operated by the Association of Universities for Research in Astronomy, Inc., under NASA 
contract NAS5-26555. BH also thanks the theoretical astrophysics group at Northwestern
University for hospitality. VK was supported in part by a Clay Fellowship at the
Harvard-Smithsonian Center for Astrophysics. FR acknowledges support
from NASA ATP grants NAG5-8460, NAG5-11396, and NAG5-12044.

\newpage
\figcaption[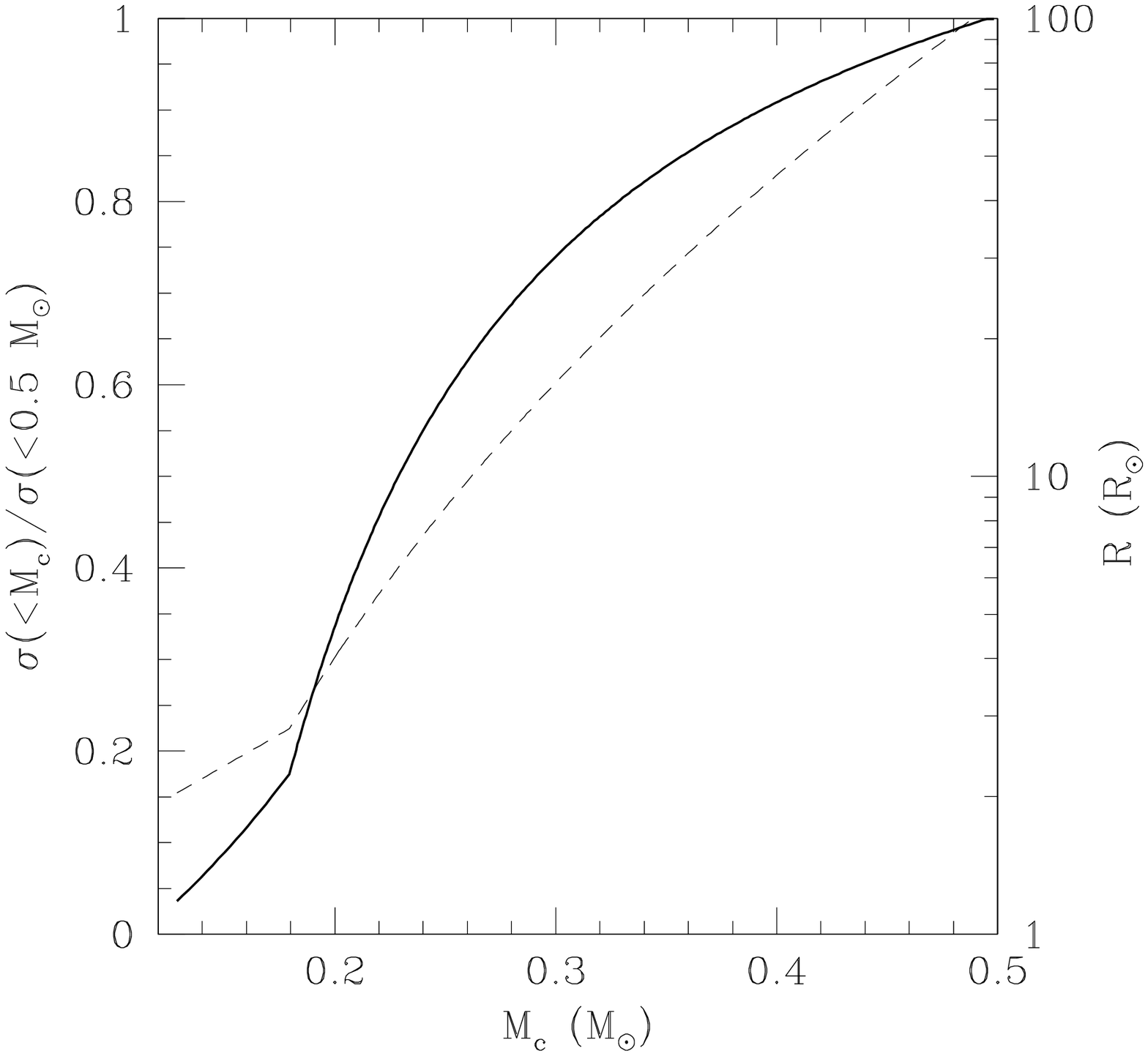]{The solid line is the cumulative cross-section for compact object-red giant collisions as a function of core mass for
 a 0.85 $\rm M_{\odot}$ star. The
dashed line indicates the radius of the giant as a function of core mass. The cross-section is dominated by the smaller core masses
(despite the smaller radii) because
the giant spends less time with a large core mass/radius.\label{dsdm}}

\figcaption[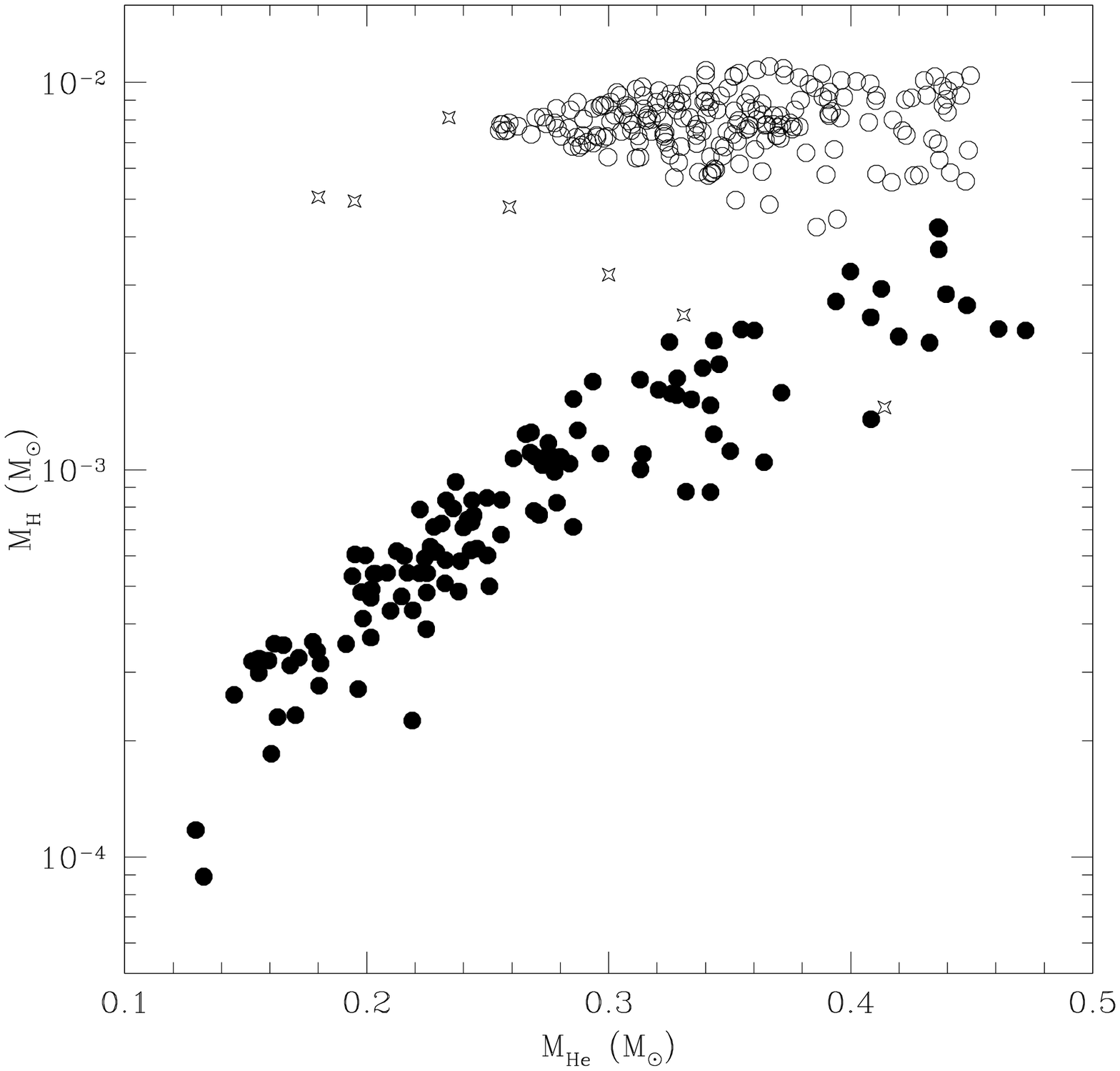]{The circles show the estimated hydrogen-layer masses for HeWDs in post-common-envelope binaries with neutron-star (open) and C/O white-dwarf (filled)
companions. The estimates have been made assuming an isentropic ($\alpha_{\rm CE}=1.5$) pre-CE envelope. The stars correspond to the layer
masses found by Driebe et al.\ (1999) in the case when the mass transfer was by stable and initiated by Roche-lobe overflow.
 \label{mh}}

\figcaption[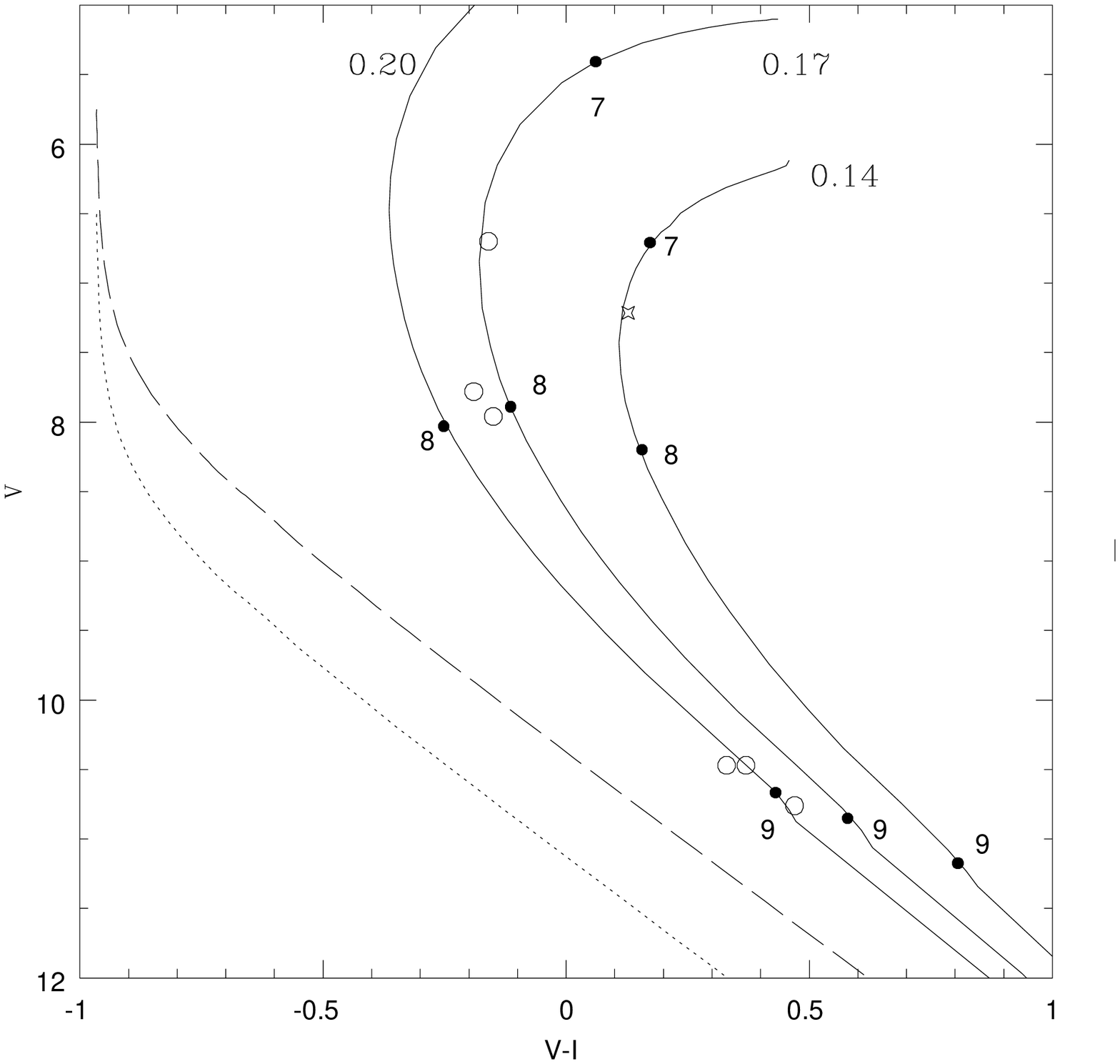]{The open circles are the data points from CGC and Taylor et al.\ (2001). The star indicates the MSP HeWD companion
in 47 Tuc (Edmonds et al.\ 2001). The filled circles
are labelled with the logarithm of the model age (in yrs). The dotted line indicates a 0.6~$\rm M_{\odot}$ C/O white dwarf
and the dashed line represents an unresolved binary of two identical such stars. The solid lines show cooling sequences for
models of 0.14, 0.17 and 0.2~$\rm M_{\odot}$ with hydrogen layer mass fraction $q_{\rm H} = 10^{-4}$. \label{LowH}}

\figcaption[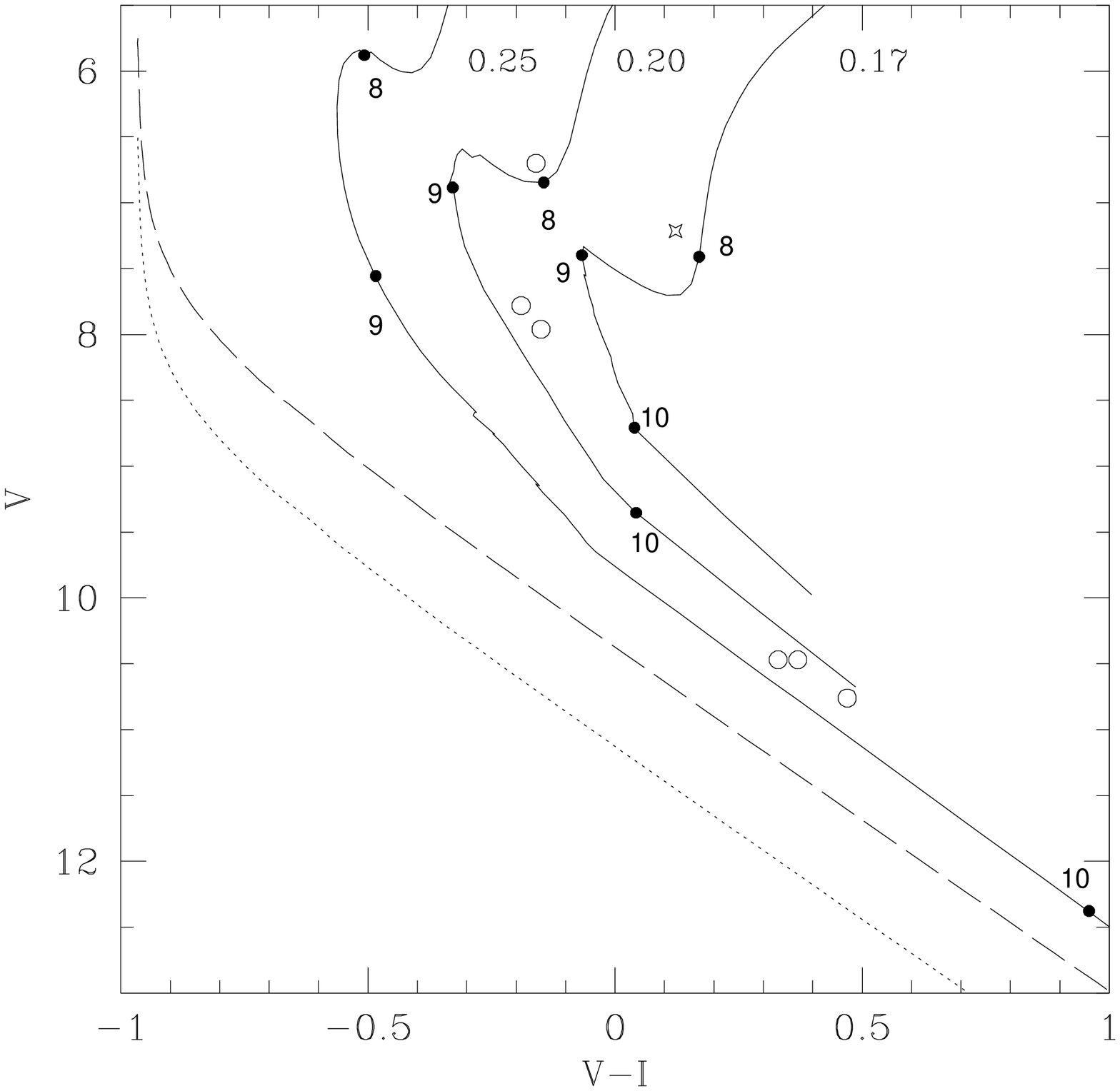]{All is as in the previous figure, although note that the three solid curves cover a
slightly different mass range (0.17-0.25$\rm \,M_{\odot}$). The initial hydrogen mass fraction is now
$q_{\rm H}=0.03$.\label{HiH}}

\figcaption[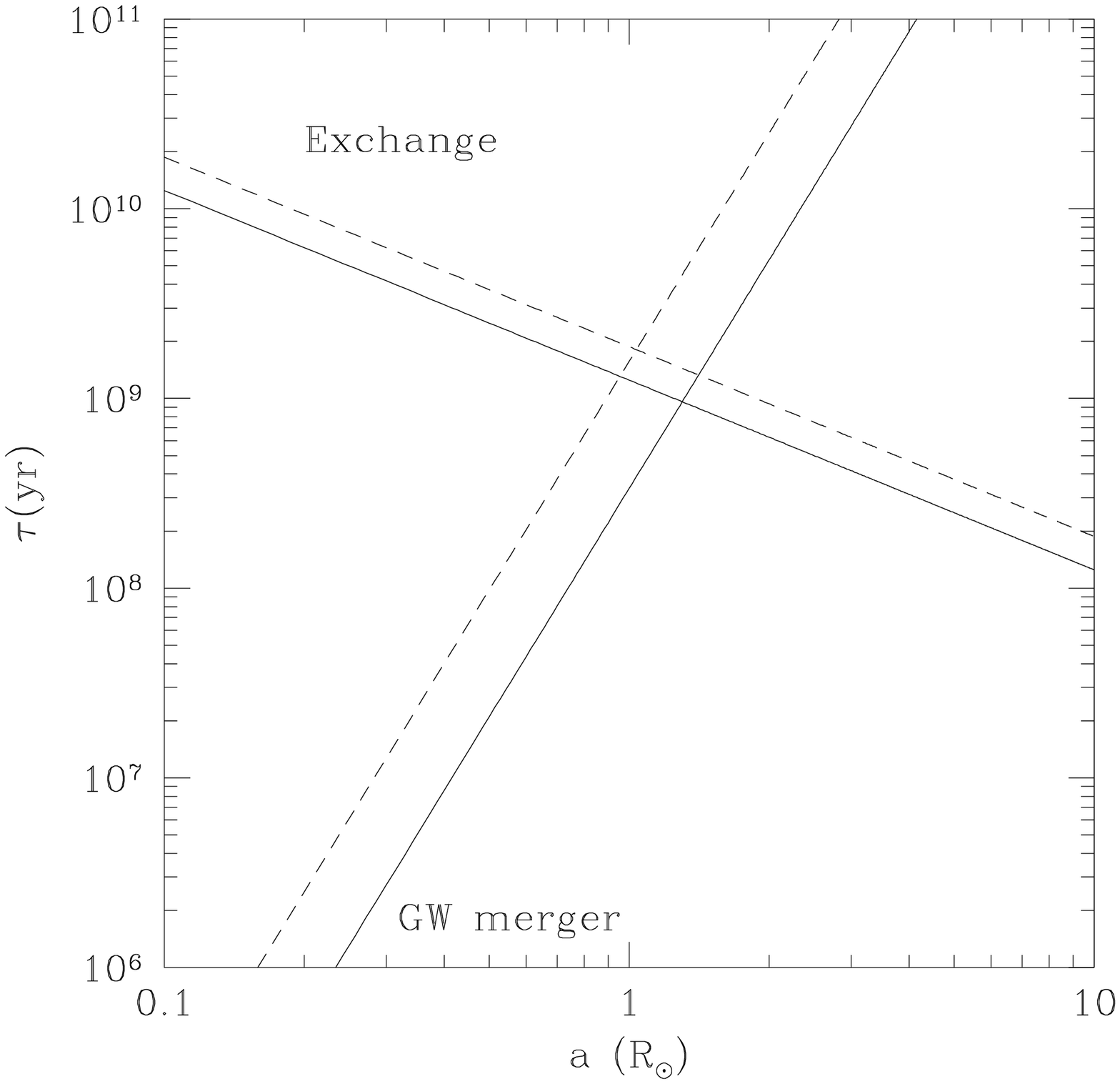]{The solid lines show the timescale for exchange and merger for a 1.4~$\rm M_{\odot}$ neutron star and a $0.2 \rm \,M_{\odot}$ white
dwarf, under the conditions appropriate for the  NGC~6397 core. The dashed lines represent the same, but for a $0.6 \rm \,M_{\odot}$ white dwarf. An additional constraint on binary
survival is that the recoil
velocity in an exchange interaction can eject a hard binary from the cluster entirely. The distribution of recoil velocities
is quite broad (see Sigurdsson \& Phinney 1993) and so is not shown on the diagram. However, binaries with separations in the range
$a \sim 1 - 10 \rm \,R_{\odot}$ have a significant chance of retention (for a cluster escape velocity of 50 km.s$^{-1}$), while
those with $a < 1 \rm \,R_{\odot}$ are frequently ejected. \label{Ta}}

\figcaption[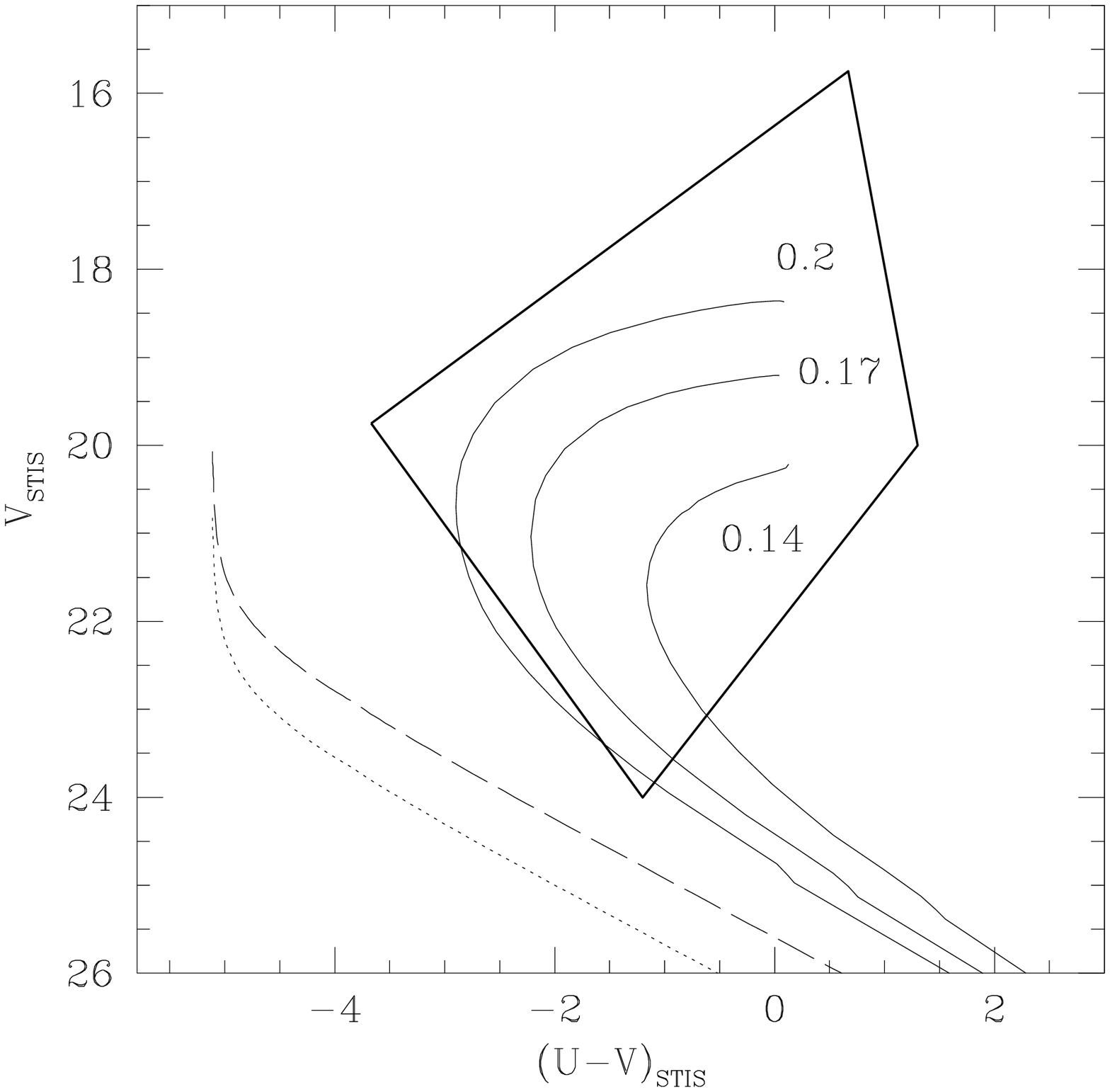]{The solid curves denote HeWD with moderate hydrogen layers. The dotted and dashed lines
indicate a $0.6 \rm \,M_{\odot}$ C/O core white dwarf and associated binary sequence. The solid box indicates the CV candidate
region defined by Knigge et al.\ (2001). Many of the objects in this region could be NFs. \label{Cool5}}

\figcaption[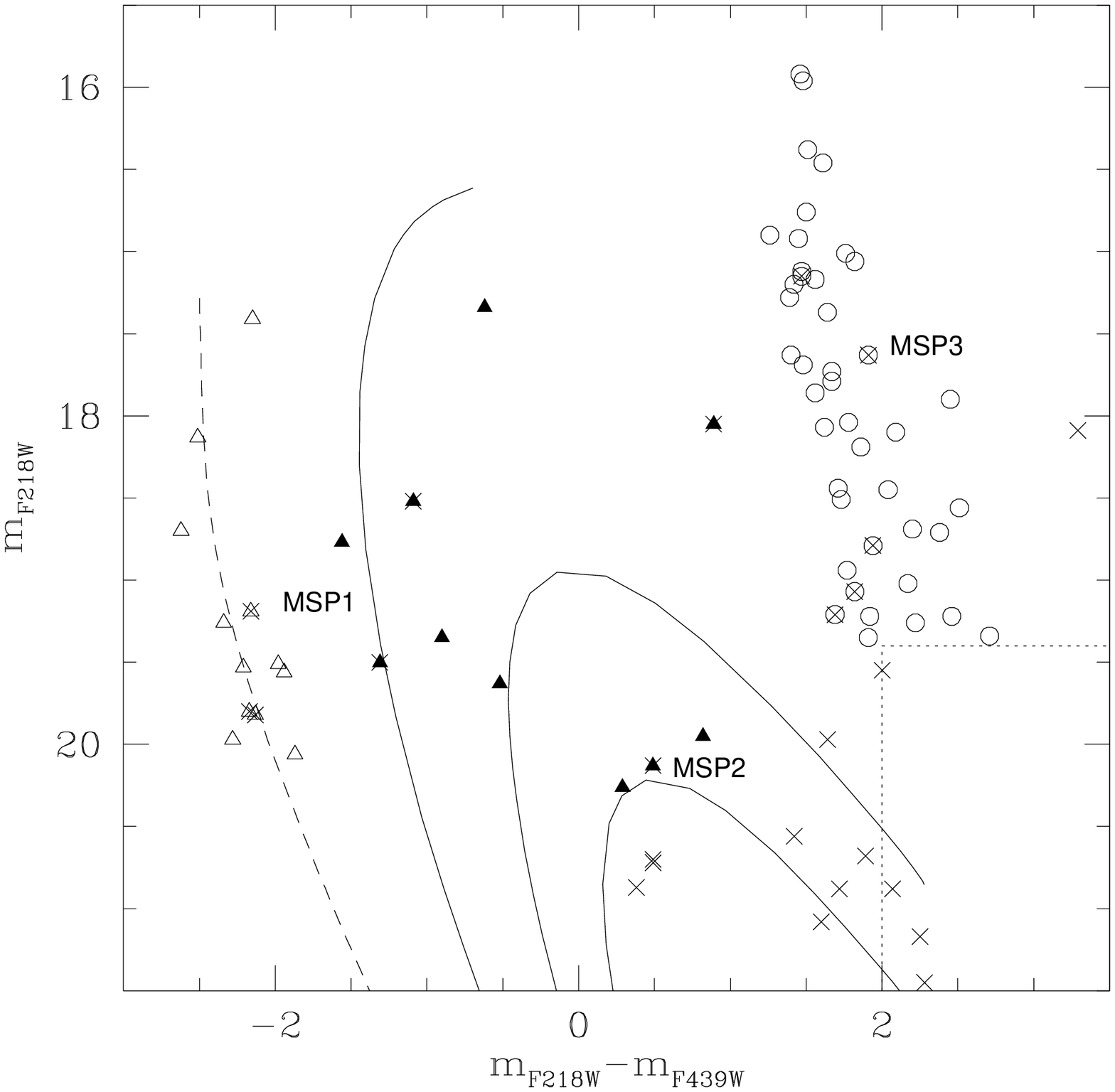]{The symbols shown are the same as in the Ferraro et al.\ (2001). Open circles and triangles denote
blue stragglers and normal white dwarfs respectively. The filled circles indicate the bright CV-candidate population. (The fainter population was not included in the tables and is thus not shown). The crosses indicate
objects with a corresponding X-ray detection. The main sequence is contained in the box in the bottom right
hand corner. The dashed line is a 0.6~$\rm M_{\odot}$ white dwarf sequence and the solid curves are for
0.3, 0.2 and 0.17~$\rm M_{\odot}$. \label{Ferr}}

\figcaption[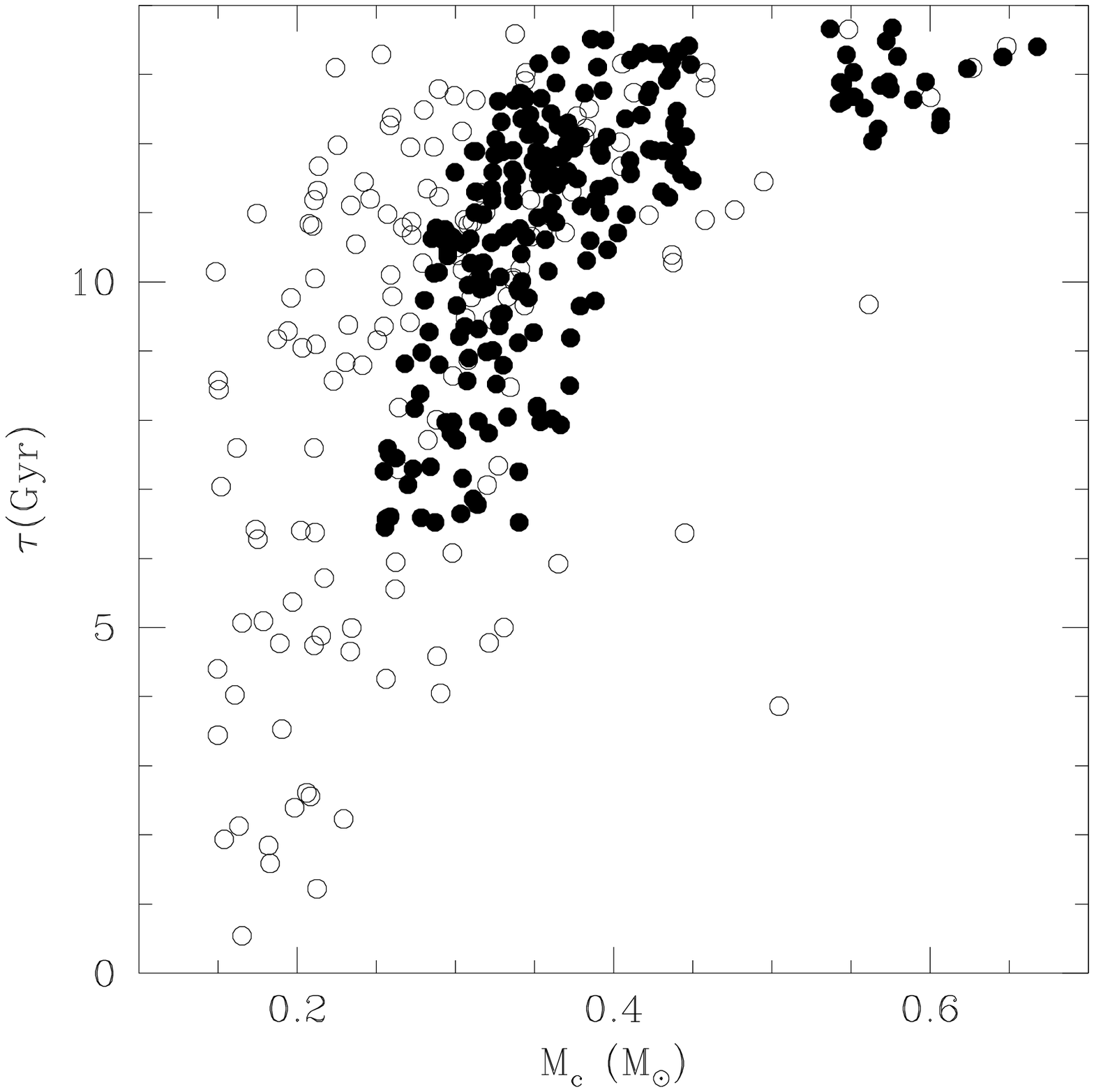]{ The ages of HeWD in binaries produced from exchange interactions in 47 Tuc, as inferred from the simulations of Rasio et al.\ (2000).
The filled circles indicate HeWDs with neutron star companions and the open circles have C/O white
dwarf companions. The age is the {\em white dwarf} age i.e. 14~Gyr minus the progenitor main sequence lifetime. The
minimum age of the HeWD with NS companions is due to the mass limit for dynamically unstable mass transfer, as
discussed in the text.
\label{MT}}

\figcaption[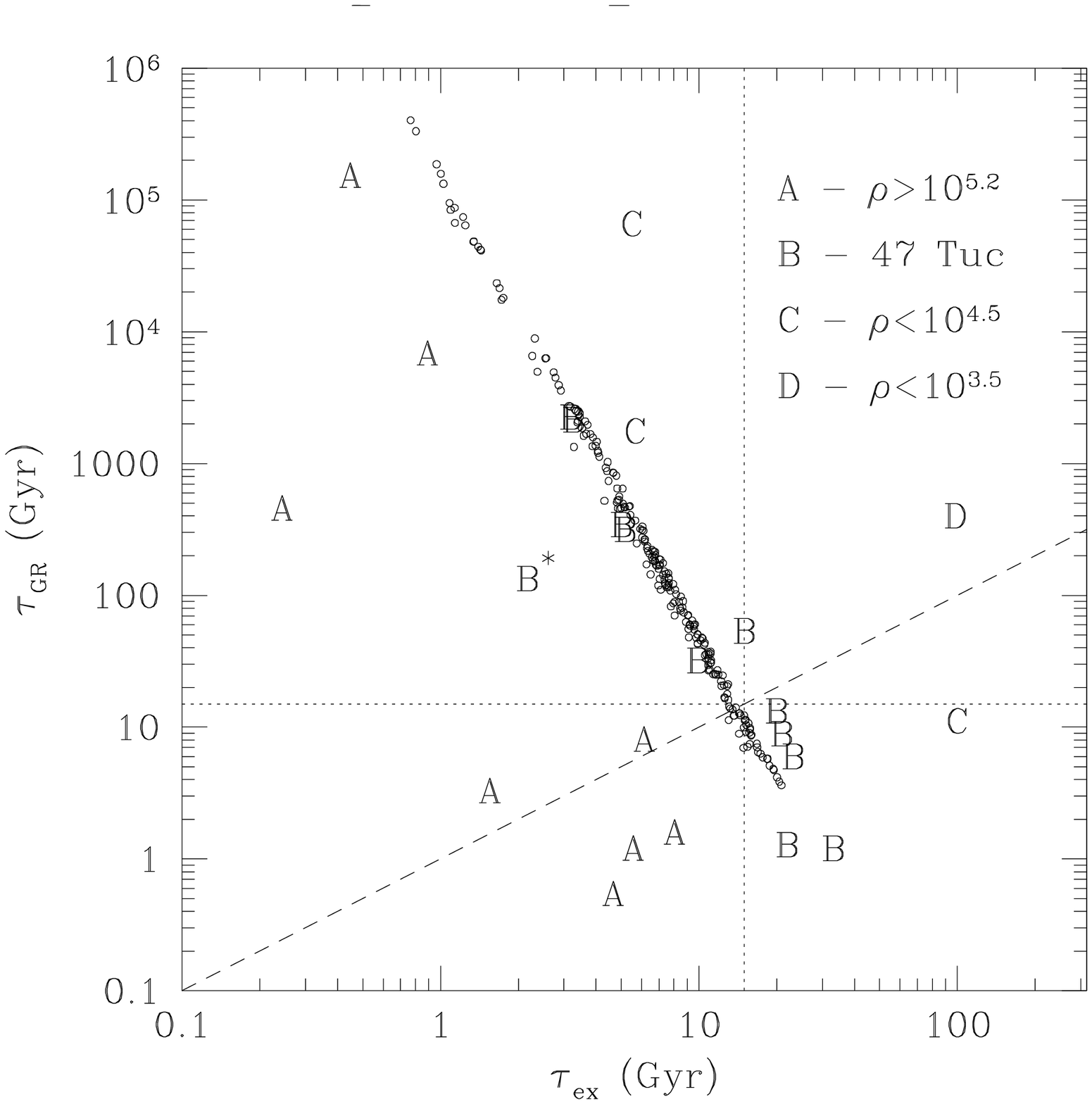]{ The disruption time and merger time for a variety of cluster binaries is shown. The classes denoted
by A, B, C and D are in order of decreasing cluster density, with B reserved for 47~Tuc alone (except for B$^*$, which
lies in NGC~6752, a cluster of very similar density. However, the orbital parameters of this system are more
characteristic of the denser clusters in class A, possibly reflecting the core-collapse nature of this cluster). The dotted lines
indicate an age of 15~Gyr. The dashed line corresponds to $\tau_{\rm m} = \tau_{\rm dis}$ and corresponds to the locus
of $\tau_{\rm max}$ as defined in equation~(\ref{Tmax}). The open circles are the results of the simulations of
the exchange interactions in 47~Tuc discussed in Rasio et al.\ (2000). The tight relation results from the fact that
the simulations produce an approximate relation between HeWD mass
 and post-common envelope orbital separation $a \sim 1 {\rm R_{\odot}} (M_{\rm He}/0.2 {\rm M_{\odot}})^{3.35}$.
\label{TT}}

\clearpage
\plotone{f1.ps}
\clearpage
\plotone{f2.ps}

\clearpage
\plotone{f3.ps}

\clearpage
\plotone{f4.ps}

\clearpage
\plotone{f5.ps}

\clearpage
\plotone{f6.ps}

\clearpage
\plotone{f7.ps}

\clearpage
\plotone{f8.ps}

\clearpage
\plotone{f9.ps}

\end{document}